\documentclass{sig-alternate}

\usepackage[utf8x]{inputenc}
\usepackage{float}
\usepackage{graphicx}
\usepackage{epstopdf}
\usepackage{epstopdf}
\usepackage{verbatim}
\usepackage{multicol}
\usepackage{multirow}
\usepackage[sort&compress,numbers]{natbib}

\usepackage[labelfont=bf]{caption}

\usepackage{color}

\newcommand{\remove}{}

\begin{document}
%\pdfrender{StrokeColor=black,TextRenderingMode=2,LineWidth=0.1pt}

%
% --- Author Metadata here ---
\conferenceinfo{submitted to WSDM}{2016}
%\CopyrightYear{2007} % Allows default copyright year (20XX) to be over-ridden - IF NEED BE.
%\crdata{0-12345-67-8/90/01}  % Allows default copyright data (0-89791-88-6/97/05) to be over-ridden - IF NEED BE.
% --- End of Author Metadata ---

\title{Geography of Emotion: Where in a City are People Happier?}
%\title{Alternate {\ttlit ACM} SIG Proceedings Paper in LaTeX
%Format\titlenote{(Produces the permission block, and
%copyright information). For use with
%SIG-ALTERNATE.CLS. Supported by ACM.}}
%\subtitle{[Extended Abstract]
%\titlenote{A full version of this paper is available as
%\textit{Author's Guide to Preparing ACM SIG Proceedings Using
%\LaTeX$2_\epsilon$\ and BibTeX} at


\numberofauthors{4} %  in this sample file, there are a *total*
\author{
% 1st. author
\alignauthor
Luciano Gallegos\\
       \affaddr{USC Information Science Institute}\\
%       \affaddr{4676 Admiralty Way}\\
       \affaddr{Marina del Rey, CA 90292}\\
       \email{\large{luciano.gallegos@gmail.com}}
%Ben Trovato\titlenote{Dr.~Trovato insisted his name be first.}\\
%       \affaddr{Institute for Clarity in Documentation}\\
%       \affaddr{1932 Wallamaloo Lane}\\
%       \affaddr{Wallamaloo, New Zealand}\\
%       \email{trovato@corporation.com}
% 2nd. author
\alignauthor
Kristina Lerman\\
       \affaddr{USC Information Science Institute}\\
%       \affaddr{4676 Admiralty Way}\\
       \affaddr{Marina del Rey, CA 90292}\\
       \email{\large{lerman@isi.edu}}
%G.K.M. Tobin\titlenote{The secretary disavows
%any knowledge of this author's actions.}\\
%       \affaddr{Institute for Clarity in Documentation}\\
%       \affaddr{P.O. Box 1212}\\
%       \affaddr{Dublin, Ohio 43017-6221}\\
%       \email{webmaster@marysville-ohio.com}
% 3rd. author
\alignauthor
Arthur Huang\\
       \affaddr{Tarleton State University}\\
       \affaddr{Stephenville, TX 76402 }\\
       \email{\large{ahuang@tarleton.edu}}
\and  % use '\and' if you need 'another row' of author names
% 4th. author
\alignauthor
David Garcia\\
       \affaddr{ETH Zurich}\\
%       \affaddr{Kreuzplatz 5}\\
       \affaddr{8092 Zurich, Switzerland}\\
       \email{\large{dgarcia@ethz.ch}}
}
%\additionalauthors{Additional authors: John Smith (The Th{\o}rv{\"a}ld Group,
%email: {\texttt{jsmith@affiliation.org}}) and Julius P.~Kumquat
%(The Kumquat Consortium, email: {\texttt{jpkumquat@consortium.net}}).}
\date{}
% Just remember to make sure that the TOTAL number of authors
% is the number that will appear on the first page PLUS the
% number that will appear in the \additionalauthors section.

\maketitle
\begin{abstract}
Location-sharing services were built upon people's desire to share their activities and locations with others.
By ``checking-in'' to a place, such as a restaurant, a park, gym, or train station, people disclose where they are, thereby providing valuable information about land use and utilization of services in urban areas. This information may, in turn, be used to design smarter, happier, more equitable cities. We use data from Foursquare location-sharing service  to identify areas within a major US metropolitan area with many check-ins, i.e., areas that people like to use. We then use data from the Twitter microblogging platform to analyze the properties of these areas. Specifically, we have extracted a large corpus of geo-tagged messages, called tweets, from a major metropolitan area and linked them US Census data through their locations. This allows us to measure the sentiment expressed in tweets that are posted from a specific area, and also use that area's demographic properties in analysis. Our results reveal that areas with many check-ins are different from other areas within the metropolitan region.
%In particular, these areas have happier tweets and residents with higher income levels, which also encourage people from other areas to commute longer distances to these places.
In particular, these areas have happier tweets, which also encourage people from other areas to commute longer distances to these places.
These findings shed light on human mobility patterns, as well as how physical environment influences human emotions.
\end{abstract}

%\category{}{Information systems}{Information systems applications}[Location based services]
\category{}{Information systems applications}{Spatial-temporal systems}[Location based services]
%A category including the fourth, optional field follows...
%\category{}{Information retrieval}{Retrieval tasks and goals}[Sentiment analysis]
\category{}{Information retrieval}{Retrieval tasks and goals}[Sentiment analysis]

\terms{Information systems}


\section{Introduction}
Happiness is an intrinsic goal of human activity. Although people have been pondering for centuries about what leads to happiness and how best to attain it, it is only in the last few decades that researchers were able to study happiness empirically. Psychologists and economists designed surveys to quantify an individual's subjective level of happiness, {subjective} well-being, or satisfaction with life, which then allowed them to study how responses to survey questions are {related to} socio-economic and demographic factors. Using such data, Easterlin~\cite{easterlin1974does} famously found that high incomes correlate with happiness.

More recently, researchers explored the geographic and environmental factors that affect happiness. They found that proximity to amenities, such as the coast or major routes of transportation, lead to higher {levels of subjective} well-being,  while proximity to a landfill negatively affects well-being~\cite{Brereton08happiness}.
Such findings offer guidelines for city planners and policy makers for designing urban areas that promote happiness and maximize equity in the distribution of resources.

One drawback of the earlier studies is that they relied on questionnaires and surveys to collect data on subjective well-being and happiness. Such data are expensive to collect, and thus limited studies to small populations, obscuring small but significant trends.
However, the rise of social media --- and location-sharing services in particular --- has given researchers an unprecedented access to geo-located data for studying the interplay between geography and happiness on a much bigger scale {and more precise temporal and geographical resolutions.
Despite these advantages, social media data sources suffer from self-selection and demographic biases~\cite{Tufekci2014}, and thus provide an alternative approach to survey data rather than a strict improvement.
It is in the combination of different methodologies that we can derive new knowledge, rather than by using each separately and arguing about which one is best.
%In this paper, we combine geo-located Twitter data with census and demographic data to exploit the potential of the combination between traditional and new data sources.

The microblogging service Twitter, for example, allows people to share short text updates with their followers and attach geographic coordinates to these posts. In addition, location-sharing services, such as Facebook Places, Gowalla, and Foursquare, allow users to simply announce their arrival to some place by ``checking-in'' to a ``venue'', which is a geo-tagged place identified by a short name. Users can either use an existing venue or create a new one. Users can even link their accounts across services, so that their Foursquare check-ins are announced to their Twitter followers or Facebook friends.
Researchers have used these publicly available location data to study human mobility~\cite{Noulas:Foursquare2011},  track people's movement~\cite{Cheng:Footprints2011}, and explore urban land use~\cite{Quercia:Deprivation2015}.

The availability of text on social media platforms also enabled researchers to analyze the sentiment of the messages and the emotional state of individuals posting them. Sentiment analysis has received much attention from the research community \cite{Kivran:Network2011}, since  it allows people to monitor sentiment on a global scale~\cite{GolderMacy} --- an impossible task if one had to rely on surveys to measure people's emotional states~\cite{Quercia:Tracking2012}.
For example, Kramer~\cite{Kramer:Unobtrusive2010} built a sentiment score
%metric between percent of words that are positive and those that are negative out of
for Facebook status updates and found that it correlates well with the self-reported satisfaction with life at the national level.%, and aggregated the metric at national US level.

Although Kramer's study suggested that one might be able to gauge a whole nation's subjective well-being and overall emotional health from the sentiment expressed on social media, it is not clear how sentiment varies on a micro-level, e.g., within communities or city neighborhoods.
%whether sentiment expressed by community residents on social media reflects community's well-being.
Eagle et al. \cite{Eagle:Network2010} showed that the  subjective well-being of communities strongly correlates with network diversity, where members of well-off communities have diverse networks while members of economically and socially disadvantaged communities have insular social relations.
%built communication networks from phone records across the entire United Kingdom, cross-referenced it with socio-economic Census data, and showed that members of well-off communities have diverse networks, while members of economically and socially disadvantaged communities have insular social relations.
%In other words, they showed that socio-economic  well-being of communities strongly correlates with network diversity.
Alshamsi et al.~\cite{Alshamsi:Misery2015} studied the effectiveness of social media in mapping happiness at finer spatial resolution and found that happy areas tend to interact with other happy areas (i.e., homophily), although they did not use other urban indicators such as demographics or mobility~\cite{Cheng:Footprints2011, Noulas:Foursquare2011}.
%Its is important noting the tricky combination of sentiment analysis and check-ins: the usage nature of location-sharing  services and the wide audience of Facebook and Twitter followers varies a lot \citep{Cramer:Performing2011}, which poses difficult in understanding patterns between human emotions and their behavior.

In this paper, we combined social media data from Twitter and Foursquare with demographic data from the US Census to carry out micro-analysis of geography and emotion. Specifically, we used Foursquare check-ins to identify areas within a major US metropolitan area that people like to use. These are the places with amenities, such as parks, restaurants, public transportation, and gyms. We then analyzed properties of these places by looking at geo-tagged messages, or tweets, coming from those areas. We linked the tweets to US Census tracts through their locations. Census tracts are small areas that are relatively homogeneous with respect to population characteristics, economic status, and living conditions. This allows us to link the sentiment expressed in tweets that are posted from different census tracts with those tracts' demographic properties.

%Our results reveal that areas with many check-ins are different from other areas within the city. In particular, these areas have happier tweets and residents with higher income levels, which also encourage people from other areas to commute longer distances to these places.
%These findings shed light on human mobility patterns, as well as how physical and environment influences human emotions that can be used to design smarter, happier, more equitable cities.

Our results reveal that areas with many check-ins are different from other areas within the city.
In particular, these areas have happier tweets, which also encourage people from other areas to commute longer distances to these places.
These findings shed light on human mobility patterns, as well as how physical and environment influences human emotions that can be used to design smarter, happier, more equitable cities.

\section{Related Works}
A number of innovative research works attempted to better understand human emotion and mobility. Some of these works focuses on geo-tagged location data extracted from Foursquare and Twitter.
Researchers reported~\cite{Cramer:Performing2011,Noulas:Foursquare2011} that Foursquare users usually check-in at venues they perceived as more interesting and express actions similar to other social media, such as Facebook and Twitter.
%On the other hand, however, researcher cannot envision all aspects of check-in data on a large scale: users share location with a wide number of audiences, which rises complexity on the number of created and chosen geo-tagged venues.
Foursquare check-ins are, in many cases, biased: while some users provide important feedback by checking-in at venues and share their engagement, others subvert the rules by deliberately creating unofficial duplicate and nonexistent venues~\cite{Duffy:Foursquare2011}.

The high availability of Foursquare and Twitter data transmitted from  mobile devices has also been subject to human mobility research~\cite{Eagle:Sensing2006}.
More specifically, some researchers used Radius of Gyration ($r_g$)~\cite{Barabasi:Rg2008} to characterize and quantify human mobility. For example, Noulas et al.~\cite{Noulas:Foursquare2011} applied $r_g$ to conduct a large-scale study of user behaviour and Foursquare check-ins with 700K users spanning a period of more than 100 days. The study revealed users' temporal and mobility patterns (the majority of users moved between 1 and 10 km and expended 100 and 2000 minutes to do so) in urban locations.
Usually, human mobility is measured at the individual's level of granularity, disclosing the users' profile as well as their personal mobility patterns, which potentially discloses information that the user may prefer to keep private~\cite{Alshamsi:Misery2015}.

Other researcher used demographic factors and associated them to sentiment analysis to measure happiness in different places.
For instance, Mitchell et al.~\cite{Mitchell:Happiness2013} generated taxonomies of US states and cities based on their similarities in word use and estimates the happiness levels of these states and cities.
Then, the authors correlated highly-resolved demographic characteristics with happiness levels and connected word choice and message length with urban characteristics such as education levels and obesity rates, showing that social media may potentially be used to estimate real-time levels and changes in population-scale measures, such as obesity rates.
Eagle et al.~\cite{Eagle:Network2010} built communication networks from phone records across the entire United Kingdom, cross-referenced it with Census data, and showed that members of well-off communities have diverse networks, while members of economically and socially disadvantaged communities have insular social relations.
Quercia et al.~\cite{Quercia:Deprivation2014, Quercia:Deprivation2015} used the Index of Multiple Deprivation (i.e., qualitative study of deprived areas in the UK local councils) to compute happiness based on small areas, providing promising fine-grained, micro-level results.
Alshamsi et al.~\cite{Alshamsi:Misery2015} studied the effectiveness of social media in mapping happiness with finer spatial resolution and, similar to \cite{Quercia:Tracking2012}, found that happy areas tend to interact with other happy areas (i.e., homophily), although other indicators such as demographic data and human mobility were not used in their research~\cite{Cheng:Footprints2011, Noulas:Foursquare2011}.

Inspired by these related works but exploring alternative approaches, we propose to use US Census tract and demographic factors to study the Los Angeles County, a large and diverse metropolitan area, and link these data with geo-tagged check-ins and tweets.
Our goal is to characterize differences between areas in Los Angeles that contain amenities, or venues that people check-in, to areas that do not. We compare the sentiment expressed in tweets that are posted from these areas and identify the physical and demographic factors that influence human emotions.
%in  of the same region, similar to the complementary publications \cite{Noulas:Foursquare2011},  \cite{Quercia:Tracking2012} and \cite{Quercia:Deprivation2014, Quercia:Deprivation2015}.

\section{Data}
\label{sec:data}
\noindent We collected a large body of tweets from Los Angeles (LA) County over the course of 4 months, starting in July 2014. Our data collection strategy was as follows. First, we used Twitter's location search API to collect tweets from an area that included Los Angeles County.  We then used Twitter4J API to collect all (timeline) tweets from users who tweeted from within this area during this time period. A portion of these tweets were geo-tagged, i.e.  they had geographic coordinates attached to them. In all, we collected 6M tweets, of which 700K made by 24K distinct users were geo-tagged.

We localized geo-tagged tweets to tracts from the 2012 US Census.\footnote{American Fact Finder} A tract is a geographic region that is defined for the purpose of taking a census of a population, containing about 4,000 residents on average, and is designed to be relatively homogeneous with respect to socio-economic characteristics of that population. We included only Los Angeles County tracts in the analysis.
A sample of the \remove{ 2010} Los Angeles Census tract map is shown in Figure~\ref{fig:map}.

\begin{figure}[h]
\centering
  \includegraphics[trim=0.5cm 1cm 1cm 2.5cm, clip=true, totalheight=0.4\textheight]{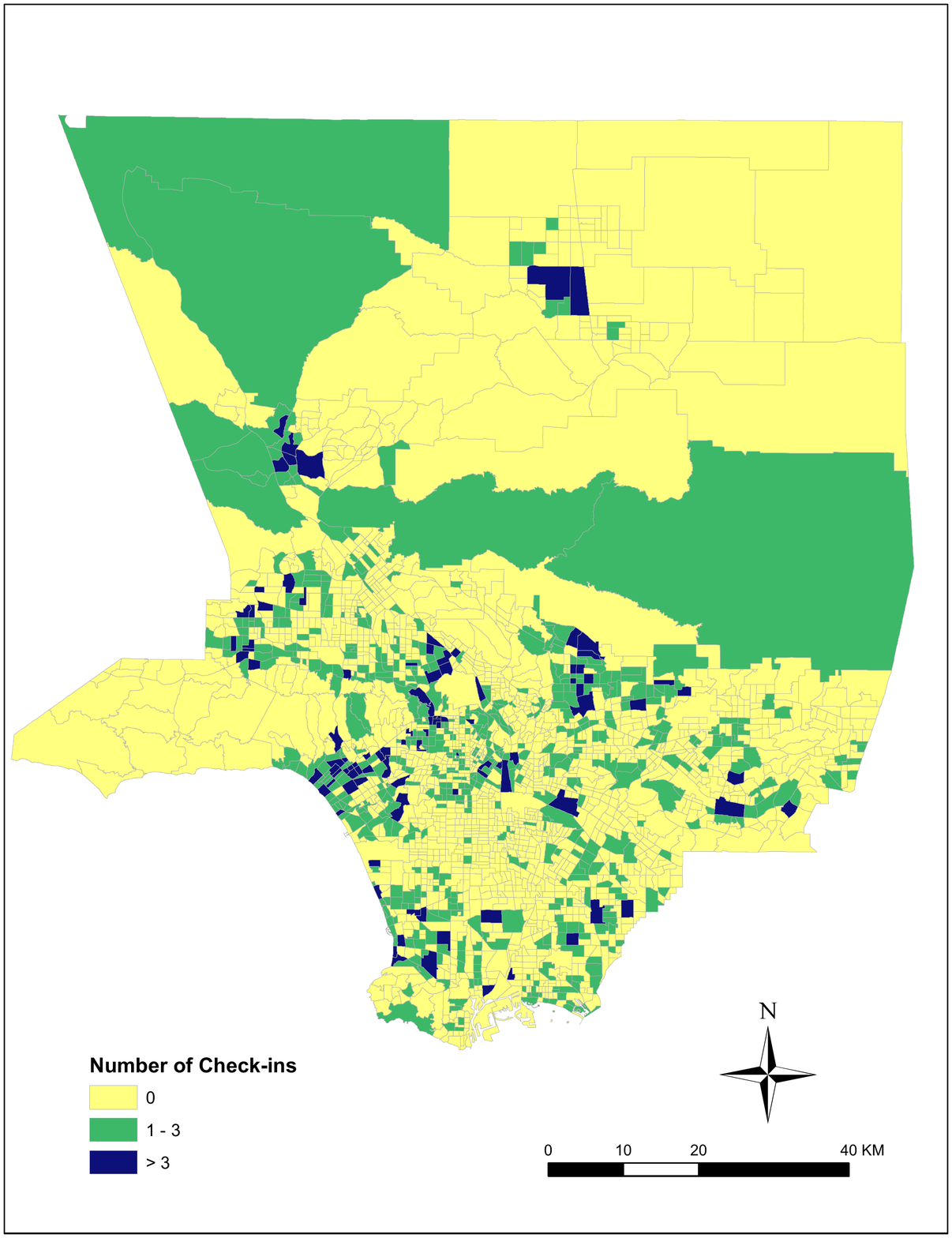}
  \caption{Los Angeles Census tract map. Tracts are colored by the number of Foursquare check-ins within them.}
  \label{fig:map}
\end{figure}

Some Foursquare users link their accounts to Twitter, so that their check-ins will be visible to their Twitter followers.  Such check-in tweets were automatically generated and had a specific format, e.g., ``I'm at 1K Studios (Burbank, CA) $http://t.co/3W5ymDM5EI$'', ``I'm at @Specialtys Cafe \& Bakery in Glendale, CA $https://t.co/IeHOY6Bbbz$'', ``I'm at Bossa Nova Brazilian Cuisine - @bossanovaeats (West Hollywood, CA) $http://t.co/pGHsMVGE3v$''. We created parsers to extract the location and venue from these tweets. In all, we extracted 5,863 check-ins from 687 tracts around Los Angeles County. The tracts in Figure~\ref{fig:map} are colored by the number of check-ins within their boundaries.

\section{Methods} \label{sec:method}
%\subsection{Sentiment analysis} \label{sec:senti}
The field of sentiment analysis~\cite{Pang2008} aims at developing tools that process text to quantify subjective states, including opinions and emotions.
Early developments in sentiment analysis extracted opinions from text, coining the term \emph{opinion mining}.
Later research focused on emotions, applying both supervised and unsupervised techniques to infer
emotional states from text.
Two recent independent surveys evaluated different sentiment analysis tools in various social media~\cite{Gonccalves2013} and in a benchmark of datasets from Twitter~\cite{Abbasi2014}.
Across social media, one of the best performing tools is SentiStrength~\cite{Thelwall2012}, which
also was shown to be the best unsupervised tool for tweets in various contexts~\cite{Abbasi2014}.

SentiStrength quantifies the  emotions expressed in text by applying a lexicon and taking into account intensifiers, negations, misspellings, idioms, and emoticons.
We apply the standard English version of SentiStrength to each tweet in our dataset, quantifying positive sentiment $P\in[+1,+5$] and negative sentiment $N\in[-1,-5$] in a way that is consistent with the Positive and
Negative Affect Schedule (PANAS) in psychology~\cite{Watson1988}.
Beyond its accuracy, SentiStrenght has been shown to perform very closely to human raters in validity tests~\cite{Thelwall2012} and has  been applied to measure emotions in  product reviews~\cite{Garcia2011}, online chatrooms~\cite{Garas2012},  Yahoo answers~\cite{Kucuktunc2012}, and Youtube comments~\cite{Garcia2012}.
In addition, SentiStrength allows our approach to be applied in the future to other languages, like Spanish~\cite{Alvarez2015}, and to include contextual factors~\cite{Thelwall2013}, like sarcasm~\cite{Rajadesingan2015}.

Research in psychology shows that emotional experiences contain components in more than two dimensions~\cite{Fontaine2007},  calling for extended analysis that includes multidimensional aspects of emotions. When measured through text, emotional meanings can be quantified through the application of the semantic
differential~\cite{Osgood1964}, a dimensional approach that quantifies emotional meaning in terms of valence, arousal, and dominance~\cite{Russell1977}. The dimension of \emph{valence} quantifies the level of pleasure expressed by a word,  \emph{arousal} measures the level of activity induced by the emotions associated with a word, and \emph{dominance}
quantifies the level of subjective power experienced during an emotion.
The state of the art in the quantification of these three dimensions is the lexicon of Warriner, Kuperman, and Brysbaert (WKB)~\cite{Warriner2013}, which includes scores in the three dimensions for almost 14,000 English lemmas.
To quantify the valence, arousal, and dominance expressed in a tweet, we lemmatize its content and apply the lexicon to compute mean values of the three dimensions as in \cite{Gonzalez2012}.
Thanks to the size of the lexicon, we find emotional terms in 82.39\% of the tweets  in our dataset, producing a multidimensional measure of emotion aggregates as expressed through tweets.

% Mobility
%\subsection{Human mobility} \label{sec:mobility}
Mobility patterns are well-correlated with demographics and individual's socio-economic status and is a current topic of academic research~\cite{Cheng:Footprints2011}.
Studies of human mobility usually focus on either the small scale (e.g., travel modes of individuals' daily commutes) or the large scale (e.g., air-travel patterns to track the spread of epidemics over time).
Some researchers adapted concepts from physics, such as the \textit{radius of gyration} ($r_g$), to characterize human mobility~\cite{Barabasi:Rg2008}. In this paper, we also use $r_g$ to quantify a Twitter user's mobility. The radius of gyration $r_g$ is the standard deviation of distances between the user's locations (given by geolocated tweets) and the center of mass of those locations. The $r_g$ measures both how frequently and how far a user moves.
A low $r_g$ indicates a user who travels mainly locally (with tweets mainly concentrated in a small geographic area), while a high radius of gyration indicates a user whose tweets are spread far apart spatially.
The $r_g$ for a user is defined as \cite{Barabasi:Rg2008}:
\begin{equation} \label{eq:rg}
	r_{g} = \sqrt{\frac{1}{n} \sum_{i=1}^{n} (r_i - r_{cm})^2}
\end{equation}
where $n$ is the number of \remove{check-ins} geotagged tweets posted by that user, and ($r_i - r_{cm}$) is the distance between a particular tweet $r_i$ and the user's center of mass $r_{cm}$. The latter is simply the average location of all tweets.

% Hypothesis testing methods 

\section{Results}
We first analyze the emotions expressed in tweets posted from different places around Los Angeles County. We find that places with many check-ins have tweets that express happier and less negative emotions.
We then explore how demographic factors contribute to these observations.

\subsection*{Sentiment analysis}
We use WKB and SentiStrength, described in Section~\ref{sec:method}, to measure the emotional content of tweets. WKB quantifies emotion along three dimensions: valence, arousal, and dominance, while SentiStrength quantifies the positive and negative sentiment expressed in tweets. Note that values measured by WKB range from 1 to 9, with 5 being the neutral score~\cite{Garcia:Positive2012}.

\begin{figure*}
  \includegraphics[width=\textwidth,height=10cm]{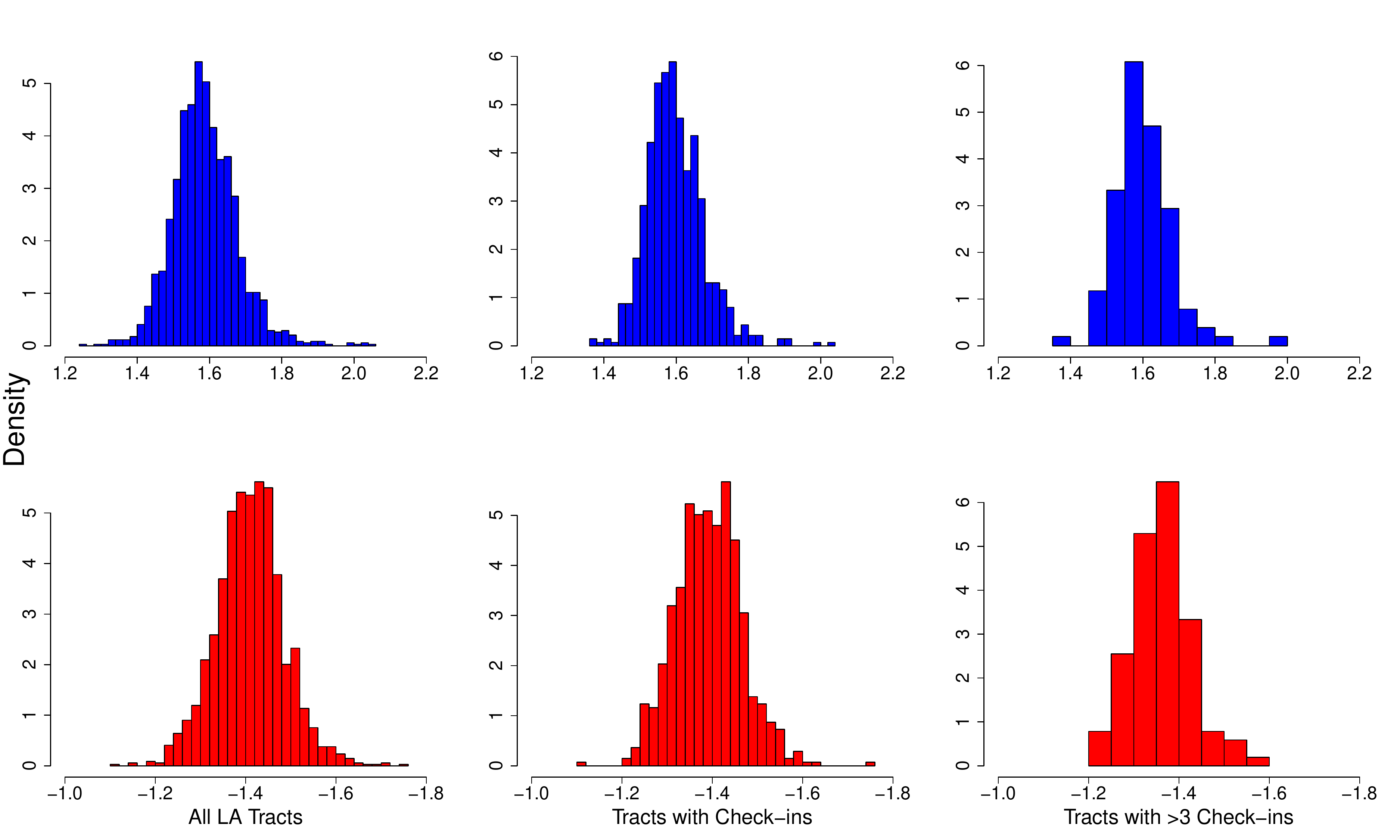}
  \caption{\textbf{SentiStrength} scores distributions by Census tracts --- Positive (top, blue) and Negative (bottom, red).}
  \label{fig:senti}
\end{figure*}

\begin{figure*}
  \includegraphics[width=\textwidth,height=10cm]{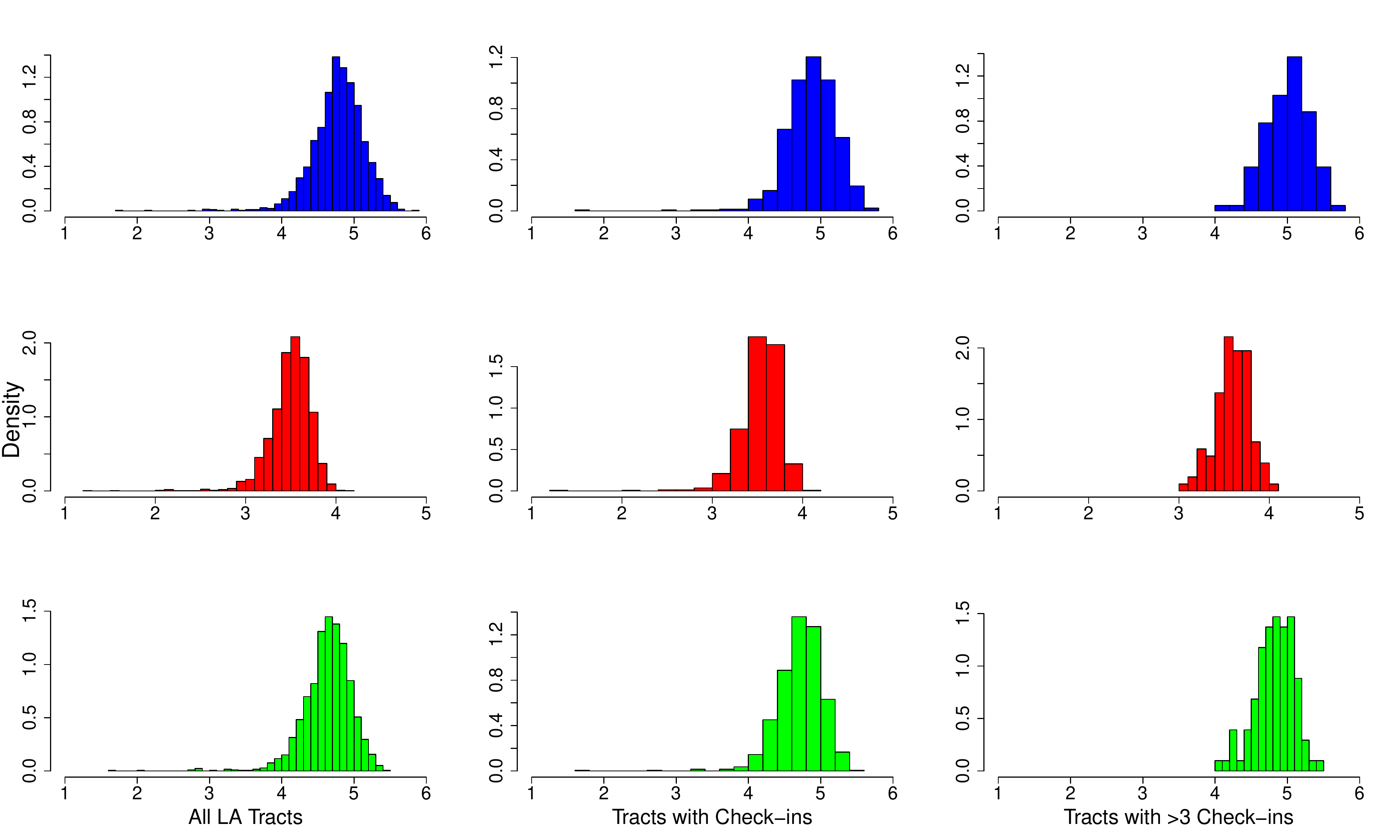}
  \caption{\textbf{WKB lexicon} scores distributions by Census tracts --- Valence (top, blue), Arousal (center, red) and Dominance (bottom, green).}
  \label{fig:warriner}
\end{figure*}

% figure here
We calculated the mean values of positive (P) and negative (N) sentiment, as well as mean valence (V), arousal (A), dominance (D) of tweets from each tract.
Figures~\ref{fig:senti} and \ref{fig:warriner} show the distribution of the means of these quantities for all 1718 tracts for which we have tweets (All LA Tracts), the 687 tracts for which we have check-ins (Tracts with Check-ins), and the 101 tracts with more than three check-ins (Tracts with >3 Check-ins).
Table~\ref{table:sentiment} reports the means of these distributions.
%Tweets from all tracts are on average slightly sad ($V<5$); however, tracts with check-ins, especially those with more check-ins, are typically happier. They also have slightly more positive and less negative sentiment than tweets from all tracts.

\begin{table*}[tbh!]
\caption{Sentiment analysis of tweets from census tracts. The first three columns give the mean values of valence, arousal and dominance of all tweets from the given set of tracts as measured using Warriner's lexicon. The last two columns and the mean values of positive and negative sentiment measured by SentiStrength from the same tweets. }
\label{table:sentiment}
\centering
	\begin{tabular}{ | c |c|| c | c | c || c | c |}
	\hline                & & \multicolumn{3}{|c||}{\textbf{WKB Lexicon}} & \multicolumn{2}{|c|}		{\textbf{SentiStrength}} \\
	\hline \textbf{Tracts} & \textbf{\# Tracts} & \textbf{Valence} & \textbf{Arousal} & \textbf{Dominance} & \textbf{Positive} & \textbf{Negative} \\
	\hline All LA Tracts        & 1718 & 4.776 & 3.493 & 4.633 & 1.588 & -1.411  \\
	\hline Tracts with Check-ins   & 687    & 4.870 & 3.534 & 4.712 & 1.598 & -1.393 \\
	\hline Tracts with >3 Check-ins & 101 & 5.001 & 3.593 & 4.825 & 1.600 & -1.362 \\
%	\hline Check-in (>5)  & 4.975 & 3.578 & 4.801 & 1.590 & -1.356 \\
	\hline
	\end{tabular}
\end{table*}

%
%The quantification mean is then calculated by each one of the 1718 Census tract within the Los Angeles county containing tweets with emotions.
%In other words, each Los Angeles county tract has its own mean of V, A and D, as well as P and N emotions.
%The \textit{emotion means} for each tract of Los Angeles allow the calculation of the means of all the Los Angeles county, as presented in the row \textit{all county} in the Table \ref{table:sentiment}.
%\note{KL: So the table represents the mean of means? Not the mean of all tweets from the tracts?}

% All check-ins
%In the analysis of results, we  first compare the different in emotions between tracts in the Los Angeles county and locations with check-in tweets.
%For this purpose, we use the same steps to calculate the means of emotion for the 687 tracts with check-in tweets, where results are presented in the row \textit{check-in} (Table \ref{table:sentiment}).
%Although check-in tracts emotion means present happier results than tracts in all county, Figures \ref{fig:senti} and \ref{fig:warriner} in \note{(KL: what's this?)} the ``Heading of Appendices'' section show that the distribution of the means are \textcolor{red}{not normally distributed, requiring a statistical test that does not assume normality.}
To determine whether the difference in the emotional content of tweets from all tracts is different from the emotional content of tweets from tracts with check-ins, we conducted the Wilcoxon rank-sum test, testing the null hypothesis that the means of these distributions are the same, and rejecting it only at confidence level of at least 95\%.
Comparing the distributions of means, we found that they are significantly different for all measurements, including mean P ($p < 0.01$), N  ($p < 10^{-5}$), V ($p < 10^{-7}$), A ($p < 0.01$), and D ($p < 10^{-6}$).
Thus, although the differences between tweets from All LA Tracts and those from Tracts with Check-ins are are small, they are significant. This difference increases further as we filter out tracts with less popular venues.
%We further filter the tweets to verify that the ``happier'' trend for check-ins holds.
%Although these statistical tests lead us to find differences with high significance
%level and reject the five null hypotheses, further variations in the check-in results
%are necessary to verify the ``happy'' trend in locations with check-in tweets.}

% Check-ins > 3
Check-in locations in Foursquare can be freely created by users, which gives rise to much noise. Users often create fake, nonsensical, and idiosyncratic locations~\cite{Cramer:Performing2011} that no one else uses. To reduce the impact of such invalid locations on our results, we filter out tracts that have only 1 or 2 checks-ins. This leaves 101 tracts with three or more check-ins (out of 697 tracts with check-ins).
%Indeed, the majority of locations have just 1 or 2 check-ins and, to avoid cheating in check-ins, we \textcolor{red}{decided} to compare differences between emotion means in the Los Angeles county and places with 3 or more check-ins tweets, the latter resulting in 101 locations out of 687 tracts with check-in tweets.
The distributions of the mean emotion scores of these tracts are shown in the last column in Figures~\ref{fig:senti} and \ref{fig:warriner}, with the means of these distributions summarized in
Table \ref{table:sentiment}. We observe that Tract with >3 Check-ins have ``happier'' (higher valence) and less negative (less negative N) tweets compared to tweets from All LA Tracts or those from Tracts with Check-ins.
%, but Figures \ref{fig:senti} and \ref{fig:warriner} in the ``Heading of Appendices'' section show that the shapes of the means of emotion are \textcolor{red}{not normal}.
Using the Wilcoxon rank-sum test, we found significant difference between the distributions of the means of N ($p < 10^{-8}$), V ($p < 10^{-7}$), A ($p < 0.001$), and D ($p < 10^{-7}$); however, we did not find a difference for P ($p < 0.1$).
%The Wilcoxon rank-sum test for means of emotions to test whether they are the same (null hypothesis) or not is then computed, considering a confidence interval of 95\%.
%The test comparing all county and check-in (>3) for SentiStrength present \textcolor{red}{no difference in P ($p < 0.1$),
%but it does for N ($p < 10^{-8}$).
%We also find significant differences for  V ($p < 10^{-7}$), A ($p < 0.001$), and D ($p < 10^{-7}$).}
These statistical tests lead us to find differences at a high significance level and thus reject the four null hypothesis above (the null hypothesis failed to be rejected only for P).
Comparing emotional content of tweets across all tracts, a trend towards ``happier'' and ``positively excited'' emotional states can be observed for tracts that have more check-ins.

%% Check-ins > 5
%The last analysis for the emotion means consists on the different between the Los Angeles county and places with 5 or more check-in tweets, described in the row \textit{check-in (>5)} in Table \ref{table:sentiment}.
%In this case, 58 locations out of 687 check-in tracts were detected.
%Due to the asymmetry in the distributions (Figures \ref{fig:senti} and \ref{fig:warriner}), the Wilcoxon rank-sum test for means of emotions with confidence interval of 95\% was computed.
%The test comparing all county and check-in (>5) for SentiStrength present different in P ($p < 0.5$) and N ($p < 10^{-6}$) emotion means.
%Difference in the emotion means also occur for the three dimension of the WKB lexicon: V ($p < 0.01$), A ($p < 0.01$), and D ($p < 0.01$).
%These statistical tests lead us to find differences with high significance level and thus reject the four null hypothesis above (only emotion means in P fail to reject the null hypothesis).
%While the trend towards happier locations can be depicted from results in all county, check-in and check-in (>3), means of emotions and statistical test do not differ between check-in (>3) and check-in (>5).
%Indeed, check-in (>3) presents better results which, suggesting that check-in (>5) starts losing data due to the empirical threshold.

\begin{figure*}[thb!]
\begin{center}
\begin{tabular} {cc}
\includegraphics[width=0.44\textwidth]{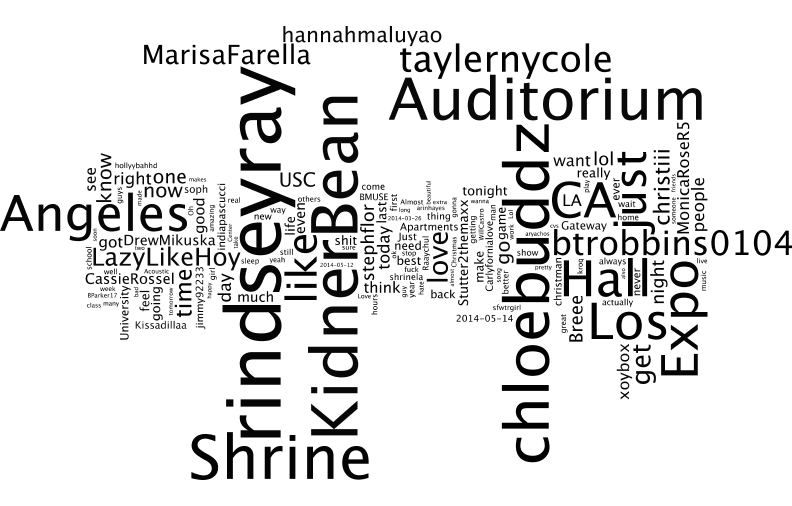} &
\includegraphics[width=0.44\textwidth]{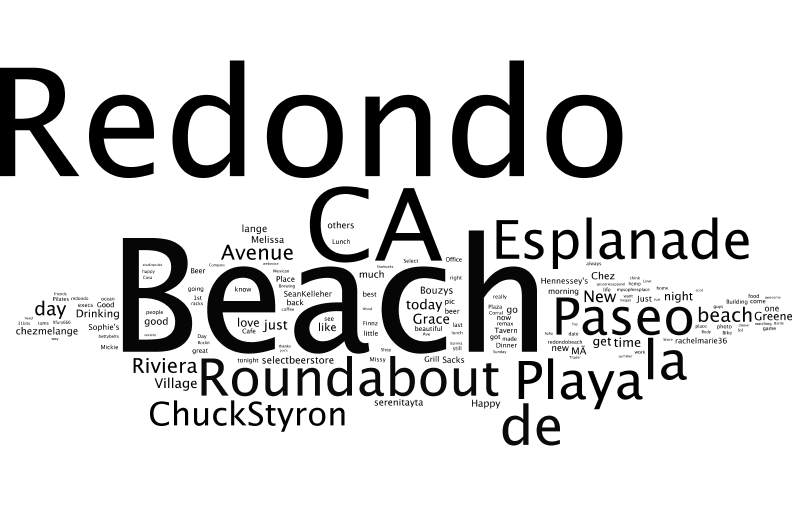} \\
(a) & (b)
\end{tabular}
\caption{Word cloud of tweets from a single tract from (a) All LA Tracts and from (b) Tracts with >3 Check-ins.}
\label{fig:wordle}
\end{center}
\end{figure*}

Why are tweets from tracts with many check-ins happier? To get insight, we look at the words that are commonly used in these tweets. Figure~\ref{fig:wordle} shows the world cloud of tweets from a single tract without check-ins and one tract that has many check-ins. The latter tract has words like ``beach'' and ``playa'' (Spanish for ``beach''), in addition to ``paseo'' and ``esplanade'', which suggest pleasant places to stroll. While tweets from the first tract have ``Shrine Auditorium'', which is a popular venue for concerts, they have fewer words associated with pleasant experiences, such as going to the beach, or strolling with friends. Though deeper analysis is required, these results suggest place with check-ins offer pleasant amenities, such as the beach, that attract people to those areas.

\subsection*{Human mobility}
\begin{figure*}
  \includegraphics[width=\textwidth,height=6.5cm]{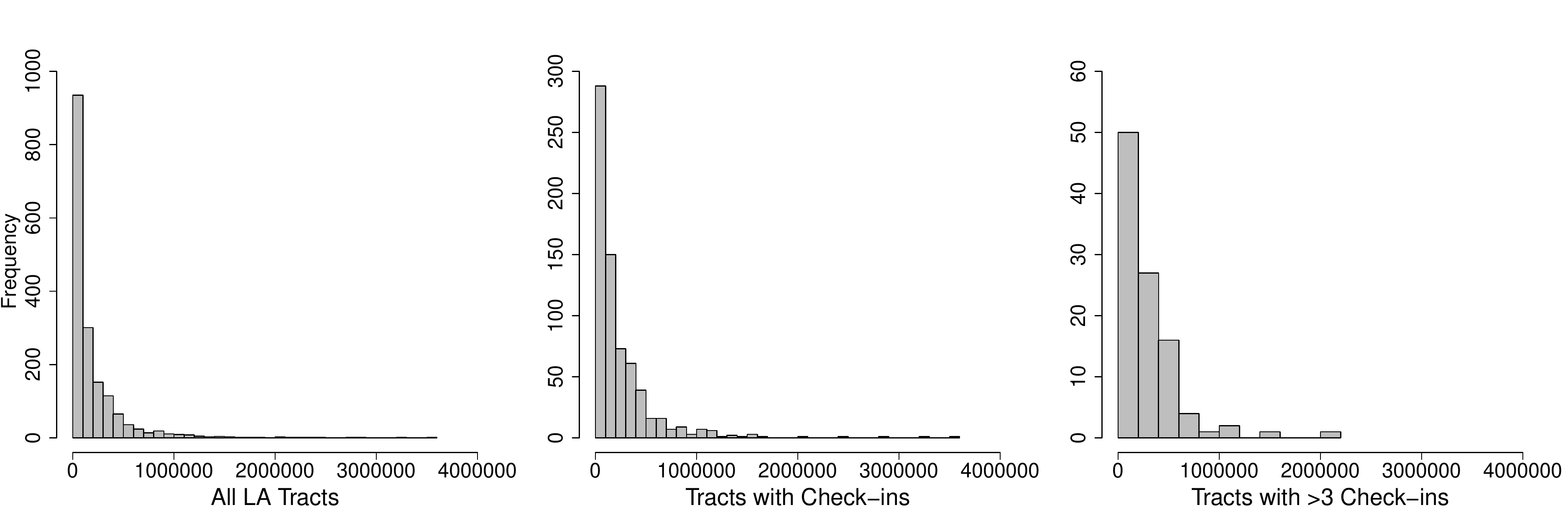}
  \caption{Distribution of user mobility by different sets of tracts. We measure mobility in a tract as average radius of gyration (in meters) of users tweeting from that tract.}
  \label{fig:rg}
\end{figure*}

% Radius of Gyration
Inspired by work such as~\cite{Barabasi:Rg2008, Cheng:Footprints2011} and as detailed in Section~\ref{sec:method}, we quantify Twitter user mobility
%means in tracts in the Los Angeles county
by computing the user's ``radius of gyration'' ($r_g$) such as in Equation~\ref{eq:rg}, which represents the average distance between locations from which the user tweets. We aggregate these values to quantify tract mobility $R_g$ as the average radius of gyration of all users tweeting from that tract:
$$R_g=\frac{1}{N}\sum_{i=1}^{N}{r_g}_i
$$
where ${r_g}_i$ is the radius of gyration of the $i$th of  $N$ users who tweet from that tract.
This allows us to compare mobility of tracts with check-ins.
Figure~\ref{fig:rg} presents the distributions $R_g$ for different sets of tracts we are comparing: all tracts in LA County with tweets, those tracts with check-ins, and tracts with >3 check-ins. Are these distributions different? The means of the distributions are: 191,925 meters for All LA Tracts, 241,261 meters for Tracts with Check-ins, and 295,057 meters  for Tracts with >3 Check-ins.
We conducted statistical inferences to test whether these distributions are the same at confidence level of 95\%.
These tests show that the means of these distributions are significantly different from each other ($p<0.001$ for all pairwise comparisons).
%These tests show that $r_g$ means are different and highly statistically significant between All LA Tracts and Tracts with Check-ins ($p < 0.001$), as well as All LA Tracts and Tracts with >3 Check-ins ($p < 0.001$), and Tracts with Check-ins and Tracts with >3 Check-ins ($p < 0.001$). From these results, we can depict that residents of check-in locations tend to travel higher $r_g$ distance and thus having higher mobility. Indeed, Tracts with >3 Check-ins residents have an $r_g$ mean about 100 km higher than All LA Tracts.

We conclude that people tweeting from tracts with many check-ins travel farther, on average, than other people in our data set. Moreover, users tweeting from tracts with more popular places (>3 check-ins) travel farthest distances. Two distinct mechanisms could explain this difference. First, people may need to travel longer distances to go to the places with more popular amenities. This would suggest that amenities that people want to use are not located equitably, forcing people to travel greater distances to use them. Alternately, however, residents of tracts with more popular amenities could have higher mobility in general (perhaps, they can better afford to travel). Although we cannot resolve between these mechanisms due to limitations of our data, doing so will have implications for city planners and land use designers.

\subsection*{Demographics}
We next examine whether differences between tracts with check-ins and those without could be explained by demographics of their residents. For this analysis, we used four demographic factors: median age, ethnicity percentage, employment percentage, and bachelor's percentage by tract.
%Figure \ref{fig:age} shows the distributions of median ages, with an average of 36 years old for all county, 37 for check-in, and 38 for check-in (>3).
Regarding age, the calculated average (median) ages are 36 years old for All LA Tracts, 37 for Tracts with Check-ins, and 38 for Tracts with >3 Check-ins.
Since the differences between median ages are small and their distribution shapes are nearly normal, we conducted statistical inferences to test whether means of these distributions are the same at confidence level of 95\%.
Median ages differences between All LA Tracts and Tracts with Check-ins are different and highly statistically significant ($p < 0.001$), as well as between All LA Tracts and Tracts with >3 Check-ins ($p < 0.001$), whereas differences between Tracts with Check-ins and Tracts with >3 Check-ins are not significant.
These results demonstrate that although the differences between median ages are small, still there is a important finding: residents in locations with check-ins and those with >3 check-ins tend to be slightly older.

%\begin{figure*}
%  \includegraphics[width=\textwidth,height=5cm]{fig-age-hist.png}
%  \caption{Age median by Census tract.}
%  \label{fig:age}
%\end{figure*}

% Ethinicity
Ethnicity is an important demographic measure.
Big cities such as Los Angeles bring together people from all around the world and a variety of ethnic groups.
For simplicity, we decided to focus on two different ethnicities: Hispanic and Non-Hispanic. Indeed, Los Angeles attracts many Hispanic groups due to its strategic position and border, which makes this city and interesting case study.
%Figures \ref{fig:hisp} and \ref{fig:nhisp} present, respectively, Hispanic and Non-Hispanic ethnicity percentages by tract.
In Non-Hispanic ethnicity tracts, the population mean is 2287 in All LA Tracts, 2744 in Tracts with Check-ins and 3379 in Tracts with >3 Check-ins.
In another direction, Hispanic ethnicity tracts present a opposite trend in the population means: 2009 in All LA Tracts,  1653 in Tracts with Check-ins and 1113 Tracts with >3 Check-ins.
Statistical inferences were also used to test whether means of these distributions are the same at confidence level of 95\%.
All null hypothesis were rejected when comparing All LA Tracts, Tracts with Check-ins and Tracts with >3 Check-ins population means for both Non-Hispanic and Hispanic ethnicities, which show the groups of populations are different with highly statistically significant results ($p < 0.001$).
%Ethnicity results show that while Non-Hispanic groups increase in Tracts with Check-ins and Tracts with >3 Check-ins locations, an opposite trend occurs with Hispanic groups.
Ethnicity results show that tracts with check-ins have high Non-Hispanic population and lower Hispanic population, suggesting that attractive amenities are located in places where fewer Hispanics live.

%\begin{figure*}
%  \includegraphics[width=\textwidth,height=5cm]{fig-hispanic-hist.png}
%  \caption{Hispanic ethnicity percentage by Census tract.}
%  \label{fig:hisp}
%\end{figure*}
%
%\begin{figure*}
%  \includegraphics[width=\textwidth,height=5cm]{fig-nhispanic-hist.png}
%  \caption{Non-Hispanic ethnicity percentage by Census tract.}
%  \label{fig:nhisp}
%\end{figure*}

% Employment
Another important demographic measure in cities is employment status of residents.
% KL - I don't really understand the sentence below.
%Since employment is a relationship between two parties (i.e., employer and employee), this measure enable us to shed light in tract's employment percentages.
%In Figure \ref{fig:employ}, employment are measured by tracts and varies between 0 to 100\%.
US Census report the percentage of  employed residents in each tract, and we use these values in our analysis. Results show that the mean employment percentage of All LA Tracts is 66\%, while Tracts with Check-ins is 67\%, and Tracts with >3 Check-ins is 69\%.
Since employment percentage distributions are nearly normal, we use the Wilcoxon rank-sum to test if these distributions are different at confidence level of at least 95\%.
While employment percentages differences between All LA Tracts and Tracts with Check-ins as well as All LA Tracts and Tracts with >3 Check-ins are statistically significant ($p < 0.001$), differences between Tracts with Check-ins and Tracts with >3 Check-ins are smaller but still significant ($p < 0.05$).
These results show that Tracts with Check-ins and Tracts with >3 Check-ins tend to have higher fraction of employed population.

%\begin{figure*}
%  \includegraphics[width=\textwidth,height=5cm]{fig-employ-hist.png}
%  \caption{Employment percentage by Census tract.}
%  \label{fig:employ}
%\end{figure*}

% Education
The last demographic measure we use is education, given by percentage of residents in a tract  who have received bachelor's degree from a college or university, a master's, professional, or doctorate degrees.
%Figure \ref{fig:bach} shows the bachelor's percentage distributions.
The bachelor's percentage means are: All LA Tracts 19\%, Tracts with Check-ins 24\%, and Tracks with >3 Check-ins (>3) 30\%.
By computing the Wilcoxon rank-sum test at confidence level of at least 95\%, we found that all null hypothesis were rejected when comparing All LA Tracts, Tracts with Check-ins and Tracts with >3 Check-ins  bachelor's percentage means with highly statistically significant results ($p < 0.001$).
These results show that Tracts with Check-ins, as well as Tracts with >3 Check-ins, are more likely to have better educated population.

%\begin{figure*}
%  \includegraphics[width=\textwidth,height=5cm]{fig-bach-hist.png}
%  \caption{Bachelor's percentage by Census tract.}
%  \label{fig:bach}
%\end{figure*}

Complementing the analysis of results of this section, we also computed correlations between these four demographic measures. Employment percentage and bachelor percentage have a positive correlation (0.39), as well as bachelor's percentage and Non-Hispanic ethnicity (0.59), and bachelor's percentage and $r_g$ (0.50).
A significant negative correlation occur between bachelor's percentage and Hispanic ethnicity (-0.75), as well as $r_g$ and Hispanic ethnicity (-0.46).
Due to the positive correlation of Non-Hispanic ethnicity and bachelor's percentage, these results also suggest that tracts %residents with bachelor's degree tend to have higher employment percentages and higher $r_g$ means.
with better educated residents (higher bachelor's) tend also to have more employed and more mobile population.
Tracts with higher percentage of Hispanic population also have less employed and less educated population, which also tends to be less mobile (lower $r_g$).
These correlations corroborate with the analysis of means and their respective statistical inferences results, as described above.

\section{Discussions and Future Works}
In this paper, we combined geo-tagged tweets, Foursquare check-ins, and demographic data from the US Census to carry out micro-analysis of geography and emotion.
Specifically, we used check-ins to identify census tracts that contain amenities that people use and publicize their use of these venues through check-ins. We then carried out sentiment analysis of the tweets posted by Twitter users from these tracts.
This allowed us to link the sentiment expressed by people in different places with the demographic properties of those places, as well as human mobility patterns.

We found that tracts with more check-in were happier places. Comparing sentiment scores of tweets from all tracts with tracts with check-ins, we observed a shift towards happier, less negative scores. We further filtered data to ignore possibly spurious or fake check-ins by removing tracts with fewer than three check-ins. We found that tracts with >3 check-in are even happier and less negative places than tracts with check-ins. However, the positive sentiment scores calculated by SentiStrength were not significantly different.
%
%Sentiment analysis presented a relationship between tracts and happier areas: Tracts with Check-ins and Tracts with >3 Check-ins show a sentiment analysis ``shift'' towards happier scores in both WKB lexicon and SentiStregth tools.
%It is remarkable to note that the difference in P found with All LA Tracts data disappears when taking into account more high quality data.
%Spurious check-ins seem to point to spam-like positive information that does not appear in other dimensions (such as for V).
This shows that including additional dimensions in analysis helps portray a more nuanced representation of emotional expression that is more robust with respect to fake content in social media.

% Risks of twitter bias, analysis - we do comparative analysis, less susceptible to bias
Our results reveal that places (tracts) with check-ins are fundamentally different from other areas within the Los Angeles County. These places offer amenities that people like to use, such as restaurants, parks, beaches, and gyms.
Indeed, population of tracts with check-ins (including >3 check-ins) tends to be slightly older, better educated, more employed, and more Non-Hispanic, compared to the population of Los Angeles County (all tracts). In addition, people tweeting from tracts with check-ins (and tracts with >3 check-ins) are more mobile, traveling farther, on average, than other people in our data set. This suggests that areas offering desirable amenities encourage people to commute longer distances to use them, although this observation may also be explained if residents of tracts with check-ins traveled more than other people.

Researchers have urged caution using when social media data, in particular Twitter, to study social science questions~\cite{Tufekci2014}. Twitter users may not be representative of the population researchers intended to study. Selection bias and other sampling effects could grossly distort the observations researchers make using Twitter data. Although we cannot eliminate all criticisms, we believe that our approach mitigates at least some of these concerns. Specifically, we conducted a \emph{comparative analysis} of user populations. Since our analysis considers differences between populations, rather than populations themselves, it is less susceptible to selection and other biases, because these will affect all populations.

Despite this, there are still limitations of our data and analysis, which prevent us from drawing important conclusions from these findings. One fundamental issue is that we are not able to distinguish between residents and visitors to the area. While demographic analysis applies to residents, we have extended it to all people tweeting from the tract. Another concern is that Foursquare users are different, perhaps they are younger and better educated, so using them to select ``attractive'' tracts may skew the data. Further work is required to address these questions. However, even with these caveats, social media offers an intriguing data source for monitoring happiness in urban areas and exploring the questions of how happiness is connected to land use.

% Challenge of analysis: distinguish between visitors and residents

Future works should consider to include socio-economic factors, such as family income by census tract, to the study of sentiments analysis and human mobility patterns.
Question on people's sociability within and between tracts, commuting preferences, as well as internet accessibility, should be considered for studying sentiments and mobility patterns in cities.
Finally, for research purposes we suggest the application of the ideas of this paper to other cities and countries.
We hope the findings described in this paper help researcher and policy maker in designing smarter, happier, more equitable cities.

%ACKNOWLEDGMENTS are optional
\section{Acknowledgments}
This paper is partially supported by the National Counsel of Technological and Scientific Development --- CNPq, Brazil.
This support is gratefully acknowledged.

%\bibliographystyle{abbrv}
%\bibliography{sigproc}  % sigproc.bib is the name of the Bibliography in this case

\balancecolumns % GM June 2007
%% That's all folks!
\end{document}